\documentstyle[sprocl]{article}

\def\Journal#1#2#3#4{{#1} {\bf #2}, #3 (#4)}

\def\NPBPS{\em Nucl. Phys. B (Proc. Suppl.)}
\def\astroph{\em Astropart. Phys.}
\def\astro{\em Astrophys. J.}
\def\NCLT{\em Let. Nuov. Cim}

\def\NPB{{\em Nucl. Phys.} B}
\def\PLB{{\em Phys. Lett.}  B}
\def\PRL{\em Phys. Rev. Lett.}

\def\be{\begin{equation}}
\def\ee{\end{equation}}
\def\bea{\begin{eqnarray}}
\def\eea{\end{eqnarray}}

\def\be{\begin{equation}}
\def\ee{\end{equation}}
\def\ba{\begin{eqnarray}}
\def\ea{\end{eqnarray}}
\def\fun#1#2{\lower3.6pt\vbox{\baselineskip0pt\lineskip.9pt
        \ialign{$\mathsurround=0pt#1\hfill##\hfil$\crcr#2\crcr\sim\crcr}}}
\def\la{\mathrel{\mathpalette\fun <}}
\def\ga{\mathrel{\mathpalette\fun >}}
\def\mpl{{{M_{{\rm Pl}}}}}

\def\prl#1#2#3{Phys. Rev. Lett. {\bf #1}, #2 (#3)}
\def\prd#1#2#3{Phys. Rev. D {\bf #1}, #2 (#3)}

\begin{document}
\begin{flushright}
CTP-TAMU-35/98\\
ACT-8/98\\
hep-ph/9809546
\end{flushright}

\vspace{12pt}

\title{CRYPTONS: A STRINGY FORM OF DECAYING SUPERHEAVY DARK MATTER, AS A
SOURCE OF THE ULTRA HIGH ENERGY COSMIC RAYS\footnote[1]{Talk
given
at the R. Arnowitt Fest: A Symposium on Supersymmetry and Gravitation,
College Station, TX; 5-7 April 1998.}}

\author{D. V. NANOPOULOS}

\address{Center for Theoretical Physics, Dept. of Physics,
Texas A \& M University,\\ College Station, TX~77843-4242, USA}
\address{Astroparticle Physics Group, HARC, Mitchell
Campus,\\ The Woodlands, TX~77381, USA}
\address{Chair of Theoretical Physics, Academy of Athens,
28~Panepistimiou~Ave., Athens~GR-10679, Greece}

\maketitle\abstracts{ Cryptons, metastable bound states of matter in the
string hidden sector, with dynamically determined masses
 $M_X\sim 10^{12}$ GeV and lifetimes $\tau_X \ga 10^{18}$ yr,
{\it may} be
generated, through inflation, with an abundance close to that
required for a near-critical universe. Their decay debris {\it may}
be
responsible for the most energetic particles striking Earth's
atmosphere. Recent developments of this {\it astonishing hypothesis} are
reviewed, indicating that NESTOR or the PIERRE AUGER project {\it may} be
able to confirm or refute the existence of cryptons.}

\section{INTRODUCTION} 
Our present understanding of the universe, as
it is encoded in the Big Bang Cosmology, suggests that there are {\it at
least} two major contributions in the energy density of the universe. 
One, is that of {\it baryonic} or {\it shiny} matter (p,n), and the other
is
that of {\it dark} matter. The rotational velocities of galaxies, the
dynamics of galaxy clusters and theories of structure formation
suggest that most of the matter in the universe is invisible and largely
composed of non-baryonic matter. The exact nature of dark matter
 is still not known and not due to lack of candidates. From massive
neutrinos to axions to the lightest supersymmetric particle (LSP)
to more exotic possibilities, we indeed have an impressive list to choose
from. While every and each of the above dark matter candidates is
plausible and originates in particle physics, the way that they form
dark matter is sharply different from the way that they form shiny
matter. One then may wonder if it is possible to replicate the way
that shiny matter is constructed, but at a higher mass scale, so that
we would have then a unified, simple explanation for the origin of
matter in the universe both shiny and dark!  Actually, as I will
discuss in this talk, such a unified explanation is possible in the
framework of string theory and {\it cryptons}~\cite{eln1}, which are
(meta)stable
bound states of matter in a hidden sector of string theory,
are the corresponding dark matter ``baryons", but superheavy, so
that when they eventually decay, their decay products may
compose \cite{eglns,BEN1} the most energetic particle striking the
Earth's atmosphere
- the ultra high energy cosmic rays (UHECR). A detailed discussion
of the data and the experimental accounts may be found in recent
reviews~\cite{WCY1}.

\section{BARYONS AND CRYPTONS}
 Let us take things from the
beginning. Let us remind ourselves how baryonic or shiny matter has
been constructed. There are two kinds of fundamental fermions,
quarks (q)  and leptons (l).  They form doublets under the
electroweak
gauge symmetry SU(2)xU(1)  e.g. (u,d) (${\nu_e},e^{-}$), while quarks
carry
fractional charges $q_u$=2/3, $q_d$=-1/3. One way to proceed from
here is to
reinvent strong interactions i.e. Quantum ChromoDynamics (QCD), in
the following way: there are no observed free fractionally-charged
 particles in nature.  That means that u and d quarks cannot float
around freely.  One then introduces a new interaction, the strong
interaction, as expressed by QCD, only for quarks, such that it
confines the quarks in color-less integer-charged particles like
pions, protons, neutrons and the likes. Actually there is a strong
correlation between the 3 of the SU(3) and the (-1/3) or (2/3) 
fractional charges as quarks form $\underline{3}$ of $SU(3)_{color}$,
and we need three
quarks to form color-singlet, integer charged baryons (p,n,...) 
which they may even be called {\it trions}. Furthermore, we may get
information about the characteristic mass scale of color-singlet
particles in the following way. The non-Abelian nature of $SU(3)_{color}$
, and the reasonable number of quark flavors, six, allow for the 
diminishing of the strong interaction coupling constant $\alpha_3(Q^2)$ as
$Q^2$
increases, thus allowing at very high energies the electroweak couplings
$\alpha_2(Q^2)$
and $\alpha_1(Q^2)$ to catch up and unify, at scale ${\cal O}(10^{16}$GeV)
as observed at LEP, in a
supersymmetric framework~\cite{EKN1}. 
       In a way, quark confinement is related to grand
unification
through its antipodal property of asymptotic freedom. Thus, given at
very high energies the common unified coupling constant,
$\alpha_{GUT}(\sim 1/25)$,
one can trace back, through standard Renormalization Group
Equations the energy scale when $\alpha_3(Q^2)$ becomes $O(1)$ i.e. it
confines. One finds that $\Lambda_{QCD} \approx {\cal O}(300$ MeV) 
as the generic characteristic mass scale
for color-singlets, in agreement with the observed hadron mass
spectrum. So that's the way that protons and neutrons are formed,
and that's how we, more or less, determine dynamically what is their
mass range. In other words, that's how baryonic or shiny matter is
constructed, the ``shiny" due, of course, to the left-over, Van der Waals
type, nuclear interactions. Actually, we know a bit more. One may prove a
simple theorem~\cite{WN1} in the perturbative framework of the Standard
Model (SM), that is of the conservation of the global Baryon (B) and
Lepton (L) numbers. In other words there has to exist a stable baryon
which happens to be the proton!  Furthermore, either
non-perturbative effects
 or beyond-standard model interactions, e.g. as exist in grand
unified theories, violate B and L quantum numbers, thus turning
proton to a metastable particle, with a lifetime $\tau_p \ga {\cal
O}(10^{33}$ yrs), rather
long-lived baryonic or shiny matter. 
        So, here are the main steps in the formation of baryonic
matter:

$\bullet$Existence of fractionally-charged particles that necessitate
the

$\bullet$Existence of a confining gauge interaction, non Abelian in
nature,
with a dynamically determined confining scale, due to the
asymptotically-free property of the non-Abelian interaction that is
partially responsible for the Grand Unification of all gauge
 coupling constants at very large energies ${\cal O}(10^{16}$ GeV) with
$\alpha_{GUT}(\sim 1/25)$, 
 that may lead to the 

$\bullet$Metastability of the lightest relevant particle and with an
appropriate
relic abundance. 

Clearly, all the above is a tall order to fulfill in the case of
dark matter, especially for the superheavy dark matter (i.e. heavier
than 1 TeV) where the danger exists that a large relic abundance
would overclose the universe. The question of abundance of
superheavy relics has recently been revisited~\cite{KRT,CKR,kuz1}.
In particular, a gravitational mechanism was
suggested~\cite{CKR,kuz1} whereby cosmological inflation may
generate a desirable abundance of such massive and weakly
interacting massive relic particles. Numerical analysis indicates
that the process may be largely independent of details of the models
considered for most properties of the dark matter constituent, as
well as of details of the transition between the inflationary phase
and the subsequent thermal radiation-dominated phase.  In the light
of this new proposal, it is interesting to revisit the possibility
that cryptons or other superheavy string relics may constitute an
important part of the astrophysical dark matter. 

\section{STRING THEORY AND CRYPTONS}
String theories have
historically been analyzed in the weak-coupling limit, where there
is an observable sector containing the known gauge interactions and
matter particles, and a hidden sector that is expected to become
strongly interacting and may play a r\^ole in supersymmetry
breaking.  In addition to the states that are massless before this
and subsequent stages of symmetry breaking, such string models also
contain Kaluza-Klein excitations with masses related to the scales
at which surplus dimensions are compactified. In the weak-coupling
limit, all these states would have masses comparable to the Planck
mass $\mpl \sim 10^{19}$~GeV, beyond the range favoured
by~\cite{CKR,kuz1}. However, the string mass estimate may be revised
downwards in the strong-coupling limit described by $M$ theory,
requiring a revised discussion as provided below. We now discuss in
more detail a specific string and $M$ theory possibility. 

{\it Heterotic-string (k=1) models:} These have been the most
studied vacua of string theory. The possibility of building explicit
models and carrying out detailed computations makes possible a
precise analysis. A well-established prediction of this class of
compactifications is the existence of light (massless at the string
scale) states which are singlets under $SU(3)_c$ and carry
fractional electrical charges, that appear generically in the hidden
gauge-group sector~\cite{FCP,FCP2}. Such particles cannot be free,
because the lightest of these particles would have to be stable and
present in the Universe with a large abundance. There are very
stringent upper limits on the abundance of such a
fractionally-charged relic, from successors of the Milliken
experiments, which are many orders of magnitude below the critical
density. However, theoretical expectations for their abundance on
Earth are about ten orders of magnitude above these
limits~\cite{FCP3}. Thus the only viable string vacua are those
where these charges are confined by a ``hidden" group $G$, as in
QCD.  The integer-charged lightest singlet bound states of such a
hidden-sector group may be stable or metastable, providing the
dark-matter candidates termed {\it cryptons}~\cite{eln1}. 

The confining group $G$ must be such that singlet bound states of
$SU(3)\times G$ have integer electric charges. For
$\displaystyle{G=\prod_N SU(N)\times\prod_n SO(2n)}$, this condition
states that~\cite{FCP2,eln2}:
\begin{equation} 
\sum_N{{i_N (N-i_N)}\over{2N}}+\sum_n \cases{0 &for $j_n =0$\cr 1/2
&for $j_n =2$\cr n/8 &for $j_n =1$\cr} 
\label{cqc} 
\end{equation}  
must be a non-vanishing integer, where for every $N$, $i_N$ is some
integer
between 0 and $N-1$. Thus the electric charge of a state
transforming
in the representation $N$ or $\overline{N}$ of $SU(N)$ and/or $2n$
of
$SO(2n)$ must be: 
\begin{equation}
 q= \pm\sum_N{i_N \over N}+\sum_n {j_n \over 2}~~{\rm mod}1, 
\end{equation}
with $\pm$ corresponding to representations $N$ or $\overline{N}$.

The case where $G$ is a product of semi-simple factors presents the
advantage, compared to a large unique semi-simple group, of
generally giving rise to a smaller number of fractionally-charged
states that have to be included in the running of the supersymmetric
Standard-Model gauge couplings.  Note also that,
because of these states, the $G$ gauge sector is not completely
``hidden". This may even be advantageous, if supersymmetry is broken
when the coupling of $G$ becomes strong, and an $F$ term is
generated.  This supersymmetry breaking would be mediated to the
observable sector not only by gravitational interactions involving
the graviton supermultiplet, but also through the usual
Standard-Model gauge interactions via the supermultiplets of
fractionally-charged states\cite{AB}. 

We now review an explicit example of a string model whose hidden
sector contains such metastable {\it crypton} bound states. This
model was originally constructed in the weak-coupling
limit~\cite{fsu}, but we expect that it may be elevated to an
authentic $M$-theory model in the strong-coupling limit.  This model
has the gauge group $SU(5)\times U(1)\times U(1)^4\times
SO(10)\times SU(4)$, with the latter two factors yielding strong
hidden-sector interactions. The following Table lists the matter
content of this hidden sector. 

\vskip 0.3cm
{\centering \begin{tabular}{|c|c|} \hline
$\Delta^0_1(0,1,6,0,-\frac{1}{2},\frac{1}{2},0)$&
$\Delta^0_2(0,1,6,-\frac{1}{2},0,\frac{1}{2},0)$\\
$\Delta^0_3(0,1,6,-\frac{1}{2},-\frac{1}{2},0,\frac{1}{2})$&
$\Delta^0_4(0,1,6,0,-\frac{1}{2},\frac{1}{2},0)$\\
$\Delta^0_5(0,1,6,\frac{1}{2},0,-\frac{1}{2},0)$& \\
$T^0_1(10,1,0,-\frac{1}{2},\frac{1}{2},0)$ &
$T^0_2(10,1,-\frac{1}{2},0,\frac{1}{2},0)$ \\
$T^0_3(10,1,-\frac{1}{2},-\frac{1}{2},0,\frac{1}{2})$ &
$T^0_4(10,1,0,\frac{1}{2},-\frac{1}{2},0)$ \\
$T^0_5(10,1,-\frac{1}{2},0,\frac{1}{2},0)$&   \\ \hline 
\end{tabular}\par}

\vspace*{0.3 cm}

{\centering \begin{tabular}{|c|c|} \hline ${\tilde
F}^{+\frac{1}{2}}_1(1,4,-\frac{1}{4},\frac{1}{4},-\frac{1}{4},\frac{1}{2})$
& ${\tilde
F^{+\frac{1}{2}}}_2(1,4,-\frac{1}{4},\frac{1}{4},-\frac{1}{4},-\frac{1}{2})$
\\ ${\tilde
F}^{-\frac{1}{2}}_3(1,4,\frac{1}{4},\frac{1}{4},-\frac{1}{4},\frac{1}{2})$
& ${\tilde
F}^{+\frac{1}{2}}_4(1,4,\frac{1}{4},-\frac{1}{4},-\frac{1}{4},6-\frac{1}{2})$
\\ ${\tilde
F}^{+\frac{1}{2}}_5(1,4,-\frac{1}{4},\frac{3}{4},-\frac{1}{4},0)$ &
${\tilde
F}^{+\frac{1}{2}}_6(1,4,-\frac{1}{4},\frac{1}{4},-\frac{1}{4},
-\frac{1}{2})$\\ ${\tilde {\bar
F}}^{-\frac{1}{2}}_1(1,4,-\frac{1}{4},\frac{1}{4},\frac{1}{4},\frac{1}{2})$
& ${\tilde {\bar
F}}^{-\frac{1}{2}}_2(1,4,-\frac{1}{4},\frac{1}{4},\frac{1}{4},-\frac{1}{2})$
\\ ${\tilde {\bar
F}}^{+\frac{1}{2}}_3(1,4,-\frac{1}{4},-\frac{1}{4},\frac{1}{4},-\frac{1}{2})$
& ${\tilde {\bar
F}}^{-\frac{1}{2}}_4(1,4,-\frac{1}{4},\frac{1}{4},\frac{1}{4},
-\frac{1}{2})$ \\ ${\tilde {\bar
F}}^{-\frac{1}{2}}_5(1,4,-\frac{3}{4},\frac{1}{4},-\frac{1}{4},0)$ &
${\tilde {\bar
F}}^{-\frac{1}{2}}_6(1,4,\frac{1}{4},-\frac{1}{4},\frac{1}{4},
-\frac{1}{2})$\\ \hline
\end{tabular}\par}

\vspace*{0.3 cm}
Table:  {\it The spectrum of hidden matter fields 
that are massless at the string
scale in the revamped flipped $SU(5)$ model.
We display the quantum numbers under the hidden 
gauge group $SO(10) \times
SO(6) \times U(1)^4$, and superscripts indicate the electric charges.}
{}\\
 
Analysis of the calculable superpotential in this model 
shows that most of these fields acquire
heavy masses just below the string scale from
couplings with singlet fields that acquire vacuum expectation values
to
cancel the $D$-term of the anomalous $U(1)$. The only light states 
that survive to have lower masses
are the $T_3, \Delta_3, {\tilde F}_{3,5}$  and
${\tilde {\bar F}}_{3,5}$. Analysis of the
renormalization-group $\beta$ functions of $SO(10)$ and $SO(6)$
suggest
that their confinement scales might lie at $\Lambda_{10}\sim
10^{14-15}$GeV for $SO(10)$ and  $\Lambda_{4}\sim 10^{11-12}$GeV for
$SU(4)$~\cite{LN1}.  This indicates that the states in the $SU(4)$
representations
$\Delta_3, {\tilde F}_{3,5}$ and ${\tilde {\bar F}}_{3,5}$ will form
the lightest bound states.

In addition to meson and baryon bound states as in QCD, 
one expects quadrilinear {\it tetron} bound states
specific to $SU(4)$~\cite{eln1}.
The mesons comprise $T_iT_j$, $\Delta_i \Delta_j$ and $
{\tilde  F_i} {\tilde {\bar F}_j}$ bound states, which are 
all short-lived, as they
decay through order $N=$ 3, 4 or 6 non-renormalizable  operators.
The
baryons have the
constituents $ {\tilde F_i} {\tilde F_j} \Delta_k$ and  $
{\tilde {\bar F}_i} {\tilde {\bar F}_j} \Delta_k$ are also
short-lived. Finally,
 there are
{\it tetrons} composed of four $\tilde F_i$s, of which the  lightest
have the forms  $ {\tilde F_i} {\tilde F_j}
{\tilde F_k} {\tilde F_l} $  and  $ {\tilde {\bar F}_i} {\tilde
{\bar
F}_j} {\tilde {\bar F}_k} {\tilde {\bar F}_l} $, where $i,j,k,l =
3,5$.
As in the case of QCD 
pions, one may expect the charged states to be
slightly
heavier than the
neutral ones, due to electromagnetic energy mass
splitting. No non-renormalizable interaction capable of enabling
this lightest bound state to decay has been 
found in a search up to eighth order.
We therefore consider that this lightest neutral {\it tetron}
is {\it a perfect candidate} for a superheavy dark matter particle. 
A rough lower bound on the lifetime of this lightest tetron is
of the order: 

\begin{equation}
\tau_X \sim  \frac{1}{M_X} {\left(  \frac{m_k}{M_X} \right) }^{10},
\end{equation}
which is very sensitive to $M_X$ and  the scale $m_k$ of suppression
of 
the non-renormalizable terms. For $m_k \sim 10^{17-18}$~GeV, 
and a tetron mass $M_X \sim 
10^{12}$GeV, we find that $\tau_X > 10^{7-17}$ years. This is a
lower
bound,
and the actual lifetime may well be considerably longer if the
leading 
decay interaction is of significantly higher order. 

\section{DARK MATTER AND CRYPTONS} 
The pressing problem we are
facing now is how does one explain why superheavy particles such as
tetrons are naturally generated with the right abundance so that to form
the
cold dark matter of the Universe? As I mentioned above
this problem was addressed by the authors in~\cite{CKR,kuz1}. They
suggested that these particles $X$ might be created through the
interaction of the vacuum with the gravitational field during the
reheating period of the Universe~\cite{bd}. Such a process involves only
the
gravitational interaction of the particle, and thus is quite
independent of the other (weak) interactions that it might have. 
This scenario leads to the following mass density of the particle
$X$ created at time $t=t_e$: 

\begin{equation}
 \Omega_X h^2 \approx \Omega_R h^2\:
\left(\frac{T_{RH}}{T_0}\right)\:
\frac{8 \pi}{3} \left(\frac{M_X}{\mpl}\right)\:
\frac{n_X(t_{e})}{\mpl H^2(t_{e})}.
\label{eq:omegachi}
\end{equation}
where  $\Omega_R h^2 \approx 4.31 \times 10^{-5}$ is the fraction of
the critical energy density that is in radiation today, and $T_{RH}$
is the 
reheating temperature.

The numerical analysis of~\cite{CKR} indicates that the correct
magnitude
for the
abundance of the $X$ particle is obtained if its mass lies in the
region $0.04 \la M_X/H \la 2$, where $H \sim 10^{13}$~GeV is the
Hubble
expansion rate at the end of inflation, which is 
expected to be of the same order as the mass of
the inflaton. For our purposes, 
we shall consider the range $10^{11}$~GeV $\la\,M_X\, \la \,
10^{14}$~GeV 
to be favourable for superheavy dark matter. It is rather remarkable
that the dynamically determined~\cite{eln1,LN1} mass range of 
the stringy cryptons, $M_X \approx {\cal O}(10^{12}$ GeV) (or more
specifically tetrons), falls just
inside the 
allowed range for superheavy particles to have naturally $\Omega_X h^2
\approx {\cal O}(1)$ !!! We thus have succeeded~\cite{eln1,BEN1} to
produce a
superheavy dark matter candidate, of purely stringy origin, that follows
exactly the same steps as sketched in section 2 that led to the
formulation of baryonic or shiny matter but now at a superheavy mass
scale. Deliberately, these steps  have been presented in a form that
applies
verbatim both for baryons and cryptons! Thus, as promised, we have
provided a realistic example where both shiny and dark matter employ
the same type of mechanism in their formation. Furthermore, back in
1991, we evaluated~\cite{eglns} the constraints on cryptons from the
possible effects of their decays on the spectrum of the microwave
background radiation and the primordially synthesized abundancies of
the light elements, from the observation of the diffuse gamma-ray
background radiation and from searches for muons and neutrinos in nuclear
decay and cosmic ray detectors. We found~\cite{eglns} that cryptons may
well have {\it the cosmological critical density} if
their lifetime exceeds $\sim 10^{16}$ yrs as imposed from the limit of
high-energy air-showers obtained by Fly's Eye atmospheric fluorescence  
detector~\cite{WCY1}. Of course, we didn't have back in 1991 a
natural mechanism of why $\Omega_X h^2    
\approx {\cal O}(1)$ as we do now, but it didn't
escape
our attention~\cite{eglns} the fact that cryptons, with their
characteristic
masses and lifetimes and of almost critical abundance, would produce
ultra-high energy cosmic rays, observable by the Fly's Eye. By a strange
twist of fate, our paper~\cite{eglns} was accepted for publication in
Nuclear 
Physics B on 22 Oct. 1991, while the
most energetic particle to strike Earth's atmosphere, ever, of
energy $(3.0 \pm 0.9)10^{20}$ eV was recorded on 15 Oct. 1991 by the Fly's
Eye!~\cite{BIR1} In
1993 the AGASA ground array detected a giant air-shower whose
primary energy was estimated to be $(1.7-2.6)10^{20}$~ eV~\cite{HY1}.
While
innocent and
benign
looking, these events have created quite a stir in the cosmic ray
community~\cite{SEE1}. Let's see why. 

\section{ULTRA HIGH ENERGY COSMIC RAYS (UHECR) AND CRYPTONS}
        
        Cosmic rays with energies up to ${\cal O}(5 \cdot 10^{18}$ eV) are
of
predominantly
galactic origin, and accept conventional explanations i.e.
acceleration by the Fermi mechanism in supernovae remnants~\cite{SG1}.
Above this energy the spectrum flattens and the composition changes
from being mostly heavy nuclei to mostly protons~\cite{WCY1}. This
correlated change in the spectrum and composition, suggests the
emergence of a population of cosmic rays from outside the Milky Way.
The Galactic magnetic field is too weak to confine protons above $10^{19}$
eV but here seems to be no credible astrophysical mechanism, either
inside or outside the Galaxy, for accelerating protons to energies above
$10^{19}$ eV! Furthermore, the most energetic cosmic-ray
particles would be affected by interactions with the ubiquitous
photons of the Cosmic Microwave Background Radiation (CMBR). If the
cosmic ray sources were far enough from us and if their energy
spectrum extended beyond, say, $10^{20}$ eV, then the ultra high energy
protons and nuclei would interact inelastically with the CMBR
photons. As shown by Greisen and Zatsepin and Kuzmin~\cite{gzk} more than
thirty years ago the
threshold for this energy degrading or sapping interaction is the onset
of pion-photon production:
\begin{equation}
E_{thres} \approx {m_p m_\pi \over E_\gamma} \approx 5\cdot 10^{19} eV
\end{equation}
for $E_{\gamma} \approx$~2 meV corresponding to $T_0=2.7^o$~K. The smooth
power law cosmic ray energy
spectrum would therefore be abruptly cut off near $5\cdot 10^{19}$~eV the
GZK
cut-off~\cite{gzk}.  The existence therefore of UHE cosmic ray events a-la
Fly's Eye~\cite{BIR1} and AGASA~\cite{HY1} discussed above create a
rather serious
problem in the following sense. Because these UHECR events go well
beyond the GZK cut-off, their sources must be within $\sim 50$~Mpc from
us,
thus by cosmological standards in the well-scrutinized local
neighborhood.
But there are few such sources so close to us and no definite
correlations have been found between the locations and arrival directions
of the most energetic events. It is worth recalling that above
$10^{19}$~eV,
the proton trajectory is so nearly straight in the Galactic and
intergalactic magnetic fields that one would have hoped, with
sufficient statistics, to pin down the sources on the celestial
sphere. So what's going on? What is the source(s) of the UHECR
events?

Clearly, one is encouraged to look~\cite{BEN1,kuz1,kuz2} for a more
exotic
or daring resolution of this basic problem, namely to identify the
UHECR events with the decay products of superheavy dark matter
particles like cryptons. In other words instead for looking for grotesque 
acceleration mechanisms of low-energy protons, let us use the
most natural source of superenergetic particles, the decay of a
superheavy particle. Indeed, as I mentioned above cryptons with 
 $\Omega_X h^2 \approx 1$
would provide~\cite{eglns} observable UHECR events at the cosmic ray
detectors 
like Fly's Eye, AGASA etc. Recently with the new developments in
string theory, where non-perturbative effects become tractable and
specifically in
the framework of the 11-dimensional M-theory, it has been a natural
reconciliation~\cite{WIT1} between the natural string scale Mstring and
the
unification scale $M_{LEP}\sim {\cal O}(10^{16}$~GeV) thus evading a severe
problem of string
phenomenology~\cite{NAN1}. This fact in conjunction with the existence of
a
natural mechanism~\cite{CKR,kuz1} discussed above providing $\Omega_X h^2
\approx 1$ for a
suitable mass range of superheavy particles triggered
 us to revisit crypton physics~\cite{BEN1}. Indeed, we did reproduce our
old results, now
standing on firmer ground, i.e. the existence of metastable, superheavy
cryptons (more accurately tetrons) of
masses ${\cal O}(10^{12}$~GeV) and lifetimes ${\cal O}(\ge 10^{18}$~yrs)
dynamically determined, and of the right
abundance  $\Omega_X h^2 \approx 1$ to be identified as the missing dark
matter. We also
argued that tetrons would be prime candidates for producing, through
their decays, the UHECR events of Fly's Eye
 and AGASA. The mass range of cryptons, $10^{12}$~GeV or $10^{21}$~eV,
and lifetime $\ge 10^{18}$~yrs seem to naively fit with the observed UHECR
events. Since such
superheavy particles would behave as cold dark matter and cluster
efficiently in all gravitational potential wells, their abundance in
our galactic halo would be enhanced above their cosmological
abundance by a factor of ${\cal O}(10^4)$. The UHECR would mainly
come from
decaying cryptons in our galactic halo, thus crossing distances
smaller than say ${\cal O}(100$~kpc), evading the GZK cut off. It is
going without
saying that the GZK cutoff is replaced by the kinematic cut off
$E \le {\cal O}(M_{crypton}\sim 10^{12}$~GeV).
A few weeks after our
paper~\cite{BEN1} appeared, Birkel and Sarkar came out
with a paper~\cite{BS1} that they were working for months. They did show, 
in very elaborated way, that cryptons are indeed ``what the doctor
ordered" as the UHECR events. They calculated the expected proton
and neutrino fluxes from decays of the superheavy metastable
particles, using the HERWIG QCD event generator. They noticed that
the predicted proton spectrum accounts for the observed spectrum of
the UHECR beyond the GZK cut-off, in shape and absolute magnitude if
the decaying superheavy particle has a mass of ${\cal O}(10^{12}$~GeV) and
a lifetime ${\cal O}(\ge 10^{20}$~yrs), if
such particles constitute all of dark matter!!! In other
words if cryptons hadn't been suggested before, Birkel and
Sarkar~\cite{BS1}
would have to invent them. They went further~\cite{BS1} and calculated
also the expected ratio of the proton to neutrino flux.  They found
that the predicted neutrino flux exceeds the proton flux for $E \le 
10^{19}$~eV and $E \ge 3 \cdot 10^{20}$~eV while the ratio $I_p \over
I_\nu$ has a characteristic peak at about $2 \cdot 10^{20}$~eV, a
rather careful diagnostic of the decaying crypton hypothesis for
future experiments such as the {\it Pierre Auger}  project. Furthermore,
since the neutrino flux dominated over the proton flux at low
energies, the bulk of the energy released by the decaying cryptons
end up as neutrinos. This should sound like music to the ears of the
NESTOR project people, because NESTOR is sensitive to neutrino fluxes of
TeV energies and the expected neutrino flux of the cryptons
decays should be something like: 
\begin{equation}
\left(\frac{10^{20} eV} {1 TeV}\right)\:
(flux)_{UCHR}\approx
10^8 \cdot \left(\frac{1 event} {km^2 \cdot century}\right)\approx
\left(\frac{10^6 events}{km^2 \cdot year}\right)
\end{equation}

        Moreover the neutrinos should be correlated in both time and
arrival direction with the cosmic rays since the path length in the
galactic halo is $\le 100$ kpc. In addition, since we are not living in
the
center of our Galaxy, but about $8.5$ kpc away, with sufficient event
statistics one should be able to discover the small anisotropy which
should result from the distribution of decaying cryptons in the
galactic halo~\cite{DT1}. 
        I believe that all the above phenomenological
analysis/predictions have 
taken cryptons out of the fancy realms of string theory and brought
 them into direct contact with the fact-based
world of Experimental Physics. We may know soon if cryptons have been
found.

\section*{Acknowledgments}
It is a great pleasure to dedicate this work to Dick Arnowitt, a
distinguished colleague and good friend.

\noindent This work was supported in part by DOE grant DE-FG03-95-ER-40917.

\end{document}